\documentclass[12pt]{article}
\usepackage{amsmath,amssymb,amscd,latexsym,euscript,mathrsfs}
\usepackage[matrix,arrow,curve]{xy}
\usepackage[cp1251]{inputenc}

\begin{document}

\begin{center}
 {\large Software for creating pictures in the \LaTeX{} environment
}
\end{center}
\centerline{
R. V. Bezhencev, anykey91@mail.ru}

\begin{abstract}
To create a text with graphic instructions for output pictures
into \LaTeX{} document, we offer software that allows us to build a picture
in WIZIWIG mode and for setting the text with these graphical instructions.

\end{abstract}


\section*{Introduction}

As we know, for drawing picture in \LaTeX{} environment, user has to
write commands, which contain itself a set of primitives, which together completed drawing.
How creating \LaTeX{}-integrated graphics and animations wrote Francesc Sunol \cite{anim}.

About drawing problems in \LaTeX{} and motivation don't integrated final image
is well described in the thesis Jie Xiao \cite{magdis}. Since the process of creating images in the \LaTeX{} environment is not a WYSIWYG (``What You See
Is what You Get''), and reduced to manual writing graphics output commands
in the \TeX{} language, the user has only to imagine how it will look finished
drawing, and approximately select control points.

This paper describes the software developed by the author of Paint\TeX{},
designed to solve this problem. It was developed in C and WinAPI, using the
methods of multi-threading, which guarantees the performance of the program.
To simplify the creation of drawings by other authors also develops software
Graphviz \cite{magdis} for drawing graphs, Drawlets \cite{magdis} for drawing
arbitrary graphics and FeynEdit \cite{feynedit} and JaxoDraw \cite{jaxodraw}
for drawing Feynman diagrams.

\section{Output line segment}

To display the line segment or vector in the user text in addition to the
coordinates of reference point, it is necessary to specify a slope angle with a width to height ratio. In the \TeX{} language command output segment is as follows:
\begin{verbatim}
\put(60,50){\line(1,-2){20}}
\end{verbatim}

where (60,50) - the coordinates of the start point of the segment, (1,-2)
- angle as the ratio of length to height, 20 - the length of the projection
on the axis $OX$. Values in a proportion of given inclination should not exceed 6 in absolute value of segments, 4 of vectors, and don't have common divisors other than 1. Details can be found in the books of \cite{lvov} and \cite{knut}.

Created by the author software Paint\TeX{} provides WYSIWYG interface for drawing images using primitives, and then converts each primitive in the appropriate command
output graphics \TeX{} language. For example, to draw the image shown in picture 1, you need a long time to calculate the coordinates of control points and other parameters
of output commands for each graphic primitive, or fit them around. This picture was painted in the program Paint \TeX{}, the output code is as follows:

\begin{verbatim}
\begin{picture}(215,283)
\qbezier(99,172)(105,172)(112,172)
\qbezier(112,172)(108,174)(105,175)
\qbezier(112,172)(108,171)(105,169)
\qbezier(63,193)(82,170)(102,147)
\qbezier(102,147)(100,151)(99,155)
\qbezier(102,147)(98,149)(95,151)
\qbezier(0,14)(111,14)(209,14)
\qbezier(209,14)(205,16)(202,17)
\qbezier(209,14)(205,13)(202,11)
\qbezier(168,22)(89,129)(8,234)
\qbezier(8,22)(93,22)(168,22)
\qbezier(8,234)(8,128)(8,22)
\qbezier(0,13)(0,145)(0,276)
\qbezier(0,276)(-1,273)(-3,269)
\qbezier(0,276)(1,273)(3,270)
\put(64,192){\circle{38}}
\put(101,160){V}
\put(8,2){O}
\put(10,283){Y}
\put(215,0){X}
\end{picture}
\end{verbatim}

\begin{center}
\begin{picture}(215,283)
\qbezier(99,172)(105,172)(112,172)
\qbezier(112,172)(108,174)(105,175)
\qbezier(112,172)(108,171)(105,169)
\qbezier(63,193)(82,170)(102,147)
\qbezier(102,147)(100,151)(99,155)
\qbezier(102,147)(98,149)(95,151)
\qbezier(0,14)(111,14)(209,14)
\qbezier(209,14)(205,16)(202,17)
\qbezier(209,14)(205,13)(202,11)
\qbezier(168,22)(89,129)(8,234)
\qbezier(8,22)(93,22)(168,22)
\qbezier(8,234)(8,128)(8,22)
\qbezier(0,13)(0,145)(0,276)
\qbezier(0,276)(-1,273)(-3,269)
\qbezier(0,276)(1,273)(3,270)
\put(64,192){\circle{38}}
\put(101,160){V}
\put(8,2){O}
\put(10,283){Y}
\put(215,0){X}
\end{picture}

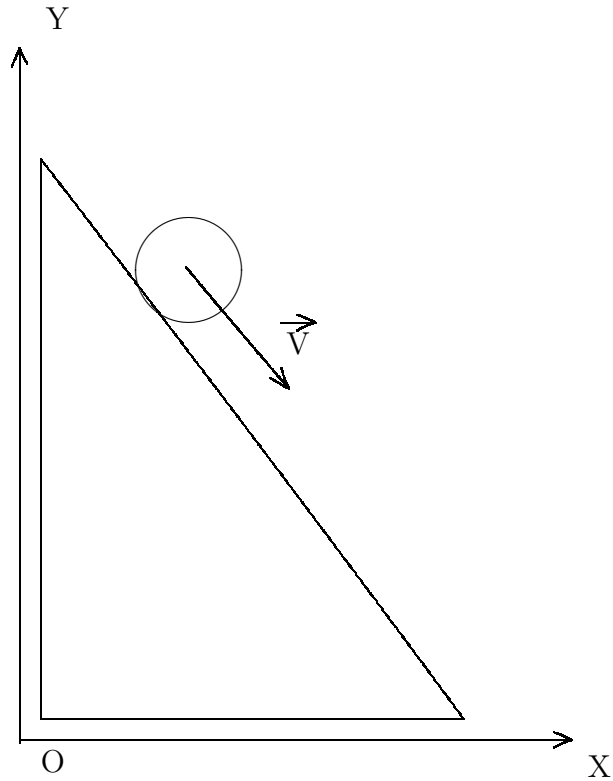
Picture 1 - Example of a picture in \LaTeX{}
\end{center}

Let us consider Paint\TeX{} in action. The user selects the desired primitive and
draws it, pointing coordinates of the reference points on which the program draws the
primitive and stores them in memory. When user save a drawing program inserts into the file
text of outputting commands of the primitive with stored coordinates of reference
points. Complex primitives are displayed in the form of a composition of simpler primitives,
for example, vector - is three straight lines, connections of the ends at the one point, which forms
the arrow, and rectangle - 4 straight. This method can display a myriad of shapes, including
three-dimensional.

\section{How the program works}

The principle of a program under development is as follows. When the program starts, a window appears
with menus, toolbar and drawing area. The user selects the primitive on the toolbar, and then sets the
coordinates of the mouse control points. Control points are stored in an instance of the class selected
shape and it draws primitive. When you select ``Save'', the program saves output commands primitives
in results file, inserting the necessary parameters (coordinates, radius) from the coordinates of the
control points.

For each primitive in the program is allocated class object of the primitive. At the
current stage of development, there are 7 classes: VETREX, LINE, LABEL, BIZE, SQVR, CIRKLE,
FISH. Each of these classes is a child of the base FIGURE class. FIGURE class content that:
\begin{verbatim}
class FIGURE
{
public:
    POINT *pt;
    FIGURE *nextFig;
	virtual void print() = 0;
};
\end{verbatim}
Due to the mechanism of inheritance, each child class inherits from a base pointer types
POINT, FIGURE and virtual function print (). When you create a primitive, start initialization
function, which converts a pointer *pt to the array points, required for a given dimension
of the primitive. That is, if you create line segment, in the constructor LINE works command
pt = new POINT[2], and if the rectangle - pt = new POINT[4] in the constructor
SQVR. Pointer *nextFig serves to form a stack of primitives. Through the mechanism of
inheritance it can point to any class of the primitive.

Each description of classes of primitives in their own redefined output function print().
This function writhen the primitive drawing commands into a text file, from which a set of
commands, and then you can copy in the article and compile \LaTeX{} tools. In each class, this
function outputs in the file own command and parameters, contained in the selected object.
Below, for example, is the content of a class of primitives ``label'':
\begin{verbatim}
class LABEL : public FIGURE
{
public:
	char *lab;
	void ini (int x, int y, char *str, int len, HDC hdc)
	{
		pt = new POINT;
		pt[0].x = x;
		pt[0].y = y;
		lab = new char[len+1];
		strcpy(lab, str);
		TextOutA(hdc, pt->x, pt->y, lab, strlen(lab));
	}
	LABEL (){}
	void print()
	{
	ofile << "\\put(" << pt[0].x - Canv_left << ","	
    << Canv_top - pt[0].y << "){" << lab << "}" << endl;
	}
}*label;
\end{verbatim}
This class contains a pointer *lab, which is converted to a string for
storing text of the label, the initialization function, which stores the
data in the structure and draws the text, the function print(), a
transformative figure in the commands of graphics output with the crop,
and a pointer *label, responsible for work stack. Function to create a
primitive ``label'' is as follows:
\begin{verbatim}
LABEL *new_label(int x, int y, char *str, int len, HDC hdc)
{
	LABEL *label_new = new LABEL;
	label_new->ini(x, y, str, len, hdc);
	if (!labelcount++) label_new->nextFig = 0;
	else label_new->nextFig = label;
	return label_new;
}
\end{verbatim}
When the user draws a primitive, in this case, the label, the function
of creating passed the coordinates to reference point, text string,
the length of the string and device handle, which will be drawn text.
Since for each new primitive memory is allocated dynamically, it have
to use for initialize the initialization function, not the designer.

When you save commands, the program for each class of primitive creates
a separate thread. Each thread runs a function that using mutex synchronizes
the output of each command. Declaration of the function follows:
\begin{verbatim}
void save(FIGURE *curfig, int counter)
\end{verbatim}
As you can see, the argument * curfig - stack pointer primitives, and
counter - their total number. Through the mechanism of inheritance,
each class primitive is a class of FIGURE, which means for synchronous
output primitives of any class is sufficient to use a single function.
So, thanks to the virtual function print (), with curfig-> print ();
You can access the output function of each primitive, and the program
will know what kind of entity it is necessary to bring in a file.

Mathematical models have been taken from the book `` Mathematical
Foundations of Computer Graphics'' \ cite {momg}. For example, a
Bezier curve - parametric curve given by the expression

$$B(t)=\sum_{i=0}^{n}P_i  b_{i,n}(t), 0<t<1$$

Where $P_i$ - function of the components of the reference peaks,
and $b_{i,n}(t)$ - basic functions of a Bezier curve, also
called the Bernstein polynomials.

$$b_{i,n}(t)=\left(n\atop i\right)t^i(1-t)^{n-i}$$

Where $\left(n\atop i\right)=\frac{n!}{I!(Ni)!}$ - Number
of combinations of $n$ on $i$, where $n$ - polynomial degree,
$i$ - number of reference peaks. Since the syntax \TeX{} can display
curves only by three points, the formula for the output has been simplified.

X = (1 - t)*(1 - t) * pt[0].x + 2*t*(1-t)*pt[1].x + t*t*pt[2].x;

Y = (1 - t)*(1 - t) * pt[0].y + 2*t*(1-t)*pt[1].y + t*t*pt[2].y;

Where pt[0].x, pt[1].x, pt[2].x - control points along the axis
of $OX$, and pt[0].y, pt[1].y, pt[2].y - coordinates of the
reference points on the axis $OY$. In the construction of the
curve, the program increments t = t + 0.01 finds points on the
curve, and then joins them small segments.

\section{The problems in during the implementation the software}

While working on the software adds the following problem. Since the values
are responsible for the slope in the primitive ``segment'' and ``vector''
must be integers, and their number is very limited, and then the slope of the
primitive there is limited number of angles. A forthcoming software user
draws segments and vectors by specifying the coordinates of the starting
and ending point. Convert their coordinates in the output instruction in
the \TeX{} language was not possible, so to print straight lines, it
was decided to use Bezier curves, defining the beginning of a line, a
middle and an end. Since the withdrawal of the Bezier curves does not specify
a value for the slope and length of the projection, the curves can be output
through the straight segments and vectors at any angle.

Just had a problem with the definition of the figure. In the \LaTeX{} drawing
area is defined manually, and the user, as well as entities that also have to
pick up some, determining what sizes will be drawing. Thanks to the automatic
cutting Paint\TeX{} defines the boundaries of the rectangle (canvas), which
was painted the image and crop a picture to fit your needs, inserting the
appropriate parameters in the command begin{picture}().

Another problem - work with coordinates Windows and \LaTeX{}. As the starting
point coordinates in Windows is the upper left edge of the window, and in the
\LaTeX{} bottom left, when converting images to files saved coordinates Windows,
and then compile the image look in the mirror image vertically. Now Paint\TeX{}
while saving the figure takes into account this nuance.

\end{document}